\documentclass{jfm}

\usepackage{amsmath,amssymb,latexsym, amscd}
\usepackage{exscale, graphics, epsfig, epstopdf, esint, color}
\usepackage{eucal, multicol}

\newcommand{\rom}[1]{\uppercase\expandafter{\romannumeral #1\relax}}

\renewcommand{\geq}{\geqslant}
\renewcommand{\leq}{\leqslant}

\DeclareMathOperator{\am}{am}

\begin{document}

\shorttitle{Stokes waves with constant vorticity}
\shortauthor{S. A. Dyachenko and V. M. Hur}
\title{Stokes waves with constant vorticity:\\ II. folds, gaps and fluid bubbles}
\author{Sergey A. Dyachenko\corresp{\email{sdyachen@math.uiuc.edu}} \and Vera Mikyoung Hur}
\affiliation{Department of Mathematics, University of Illinois at Urbana-Champaign\\ Urbana, IL 61801 USA}

\maketitle

\begin{abstract}
The Stokes wave problem in a constant vorticity flow is formulated, by virtue of conformal mapping techniques, as a nonlinear pseudodifferential equation, involving the periodic Hilbert transform, which becomes the Babenko equation in the irrotational flow setting. The associated linearized operator is self-adjoint, whereby the modified Babenko equation is efficiently solved by means of the Newton-Conjugate Gradient method. 

For strong positive vorticity, a `fold' appears in the wave speed versus amplitude plane, and a `gap' as the vorticity strength increases, bounded by two touching waves, whose profile contacts with itself at the trough line, enclosing a bubble of air. More folds and gaps follow for stronger vorticity. 

Touching waves at the beginnings of the lowest gaps tend to the limiting Crapper wave as the vorticity strength increases indefinitely, while the profile encloses a circular bubble of fluid in rigid body rotation at the ends of the gaps. Touching waves at the beginnings of the second gaps tend to the circular vortex wave on top of the limiting Crapper wave in the infinite vorticity limit, and the circular vortex wave on top of itself at the ends of the gaps. Touching waves for higher gaps accommodate more circular bubbles of fluid. 
\end{abstract}


\section{Introduction}\label{sec:intro}

\cite{Stokes1847, Stokes1880} made formal but far-reaching considerations about periodic waves at the surface of an incompressible inviscid fluid in two dimensions, subject to the force of gravity, propagating a long distance at a practically constant velocity without change of form. For instance, he observed that the crests become sharper and the troughs flatter as the amplitude increases, and that the wave of greatest height, or the extreme wave, is distinguished by a $120^\circ$ peaking at the crest. It would be impossible to give here a complete account of Stokes waves. We merely pause to remark that in the irrotational flow setting, notable recent advances were based on a formulation of the problem as a nonlinear pseudodifferential equation, due to \cite{Babenko1987} and others. For instance, \cite{DLK2016, Lushnikov2016, LDS2017} revealed the complex singularities of almost extreme waves in meticulous detail. 

The irrotational flow assumption is well justified in some situations. But rotational effects are significant in many others. Examples include wind driven waves, waves in a shear flow, and waves near a ship or pier. Constant vorticity is of particular interest. For instance, when waves are short compared with the characteristic lengthscale of vorticity, it is the value of the vorticity at the fluid surface which influences the waves \citep[see][for instance]{PTdS1988}. Moreover, constant vorticity is representative of a wide range of physical scenarios. Examples include tidal currents, the alternating lateral movement of water in conjunction with the rise and fall of the tide, for which positive and negative constant vorticity are appropriate for the ebb and flood. By the way, here we distinguish negative vorticity for waves propagating downstream, and positive vorticity for upstream. 

\cite{DH1} modified the Babenko equation and supplemented it with a scalar constraint, to permit constant vorticity and finite depth, and numerically solved by means of the Newton-GMRES method. Here we manipulate a change of variables to eliminate the Bernoulli constant from the modified Babenko equation and, hence, the scalar constraint. The associated linearized operator becomes self-adjoint, whereby efficiently handled by means of the Conjugate Gradient method. As a result, we achieve a high numerical precision for large amplitude waves for unprecedentedly large values of constant vorticity. 

\cite{SS1985} discovered, and \cite{DH1} rediscovered, that for a large value of positive constant vorticity, a fold appears in the wave speed versus amplitude plane, and a gap of unphysical solutions as the value of the vorticity increases, bounded by two touching waves, whose profile contacts with itself at the trough line, enclosing a bubble of air. Here we find {\em more folds and gaps} as the value of the vorticity increases further. See Figure~\ref{fig:c-s}, for instance, for five folds and five gaps! 

Moreover, \cite{DH1} discovered that touching waves at the beginnings of the lowest gaps tend to the limiting Crapper wave \citep[see][for instance]{Crapper1957} as the value of positive constant vorticity increases indefinitely, revealing a striking and surprising link between rotational and capillary effects, while the profile encloses a circular bubble of fluid in rigid body rotation at the ends of the gaps. Here we find the `circular vortex wave' on top of the limiting Crapper wave at the beginning of the second gap in the infinite vorticity limit, and the circular vortex wave on top of itself at the end of the gap. We find {\em more circular bubbles of fluid} at the boundaries of higher gaps in the infinite vorticity limit. See Figure~\ref{fig:5c}, for instance, for a profile nearly enclosing five circular bubbles of fluid!

\section{Formulation}\label{sec:formulation}

We consider a two dimensional, infinitely deep and constant vorticity flow of an incompressible inviscid fluid, subject to the force of gravity, and the wave motion at the interface between the fluid and air. We assume for simplicity the unit density. Suppose that in Cartesian coordinates, the $x$ axis points in the direction of wave propagation, and the $y$ axis vertically upward. Suppose that the fluid occupies a region $\varOmega(t)$ in the $(x,y)$ plane at time $t$, bounded above by a free surface $\varGamma(t)$. 

Let $\boldsymbol{u}=\boldsymbol{u}(x,y,t)$ denote the velocity of the fluid at the point $(x,y)$ and time $t$, and $P=P(x,y,t)$ the pressure. They satisfy the Euler equations for an incompressible fluid:
\begin{equation}\label{E:Euler}
\boldsymbol{u}_t+(\boldsymbol{u}\cdot\nabla)\boldsymbol{u}=-\nabla P+(0,-g)\quad\text{and}\quad
\nabla\cdot\boldsymbol{u}=0\quad\text{in $\varOmega(t)$,}
\end{equation}
where $g$ is the constant of gravitationals acceleration. Throughout, we assume that the vorticity
\[
\omega=\nabla\times\boldsymbol{u}
\] 
is constant. 
We assume that there is no motion in the air and we neglect the effects of surface tension. The kinematic and dynamic conditions state that
\begin{equation}\label{E:surface}
\partial_t+\boldsymbol{u}\cdot\nabla\text{ is tangent to $\bigcup_t\varGamma(t)$}
\quad\text{and}\quad P=P_{atm}\text{ at $\varGamma(t)$,}
\end{equation}
respectively, where $P_{atm}$ is the constant atmospheric pressure. 
We assume without loss of generality that $\varOmega$ and $\boldsymbol{u}$, $P$ are $2\pi$ periodic in the $x$ variable.

Suppose that
\begin{equation}\label{def:phi}
\boldsymbol{u}(x,y,t)=(-\omega y,0)+\nabla\phi(x,y,t)
\end{equation}
for $(x,y)\in\varOmega(t)$, whence 
\begin{equation}\label{E:phi}
\Delta\phi=0 \quad\text{in $\varOmega(t)$} 
\end{equation}
by the latter equation of \eqref{E:Euler}. We merely pause to remark that for any $\omega\in\mathbb{R}$,
\[
\varOmega(t)=\{(x,y)\in\mathbb{R}^2:y<0\},\quad\boldsymbol{u}(x,y,t)=(-\omega y,0)
\quad\text{and}\quad P(x,y,t)=P_{atm}-gy
\]
solve \eqref{E:Euler} and \eqref{E:surface} at all times. Here we assume that some external effects produce vorticity and we restrict the attention to waves propagating in a perturbed rotational flow. For nonconstant vorticity, 
\eqref{def:phi} and \eqref{E:phi} are inapplicable. 

Substituting \eqref{def:phi} into the former equation of \eqref{E:Euler} and using the latter equation of \eqref{E:surface}, we make an explicit calculation to arrive at 
\begin{equation}\label{E:bernoulli}
\phi_t+\frac12|\nabla\phi|^2-\omega y\phi_x+\omega\psi+gy=b(t)
\quad\text{at $\varGamma(t)$}
\end{equation}
for some function $b$, where $\psi$ is a harmonic conjugate of $\phi$. By the way, since $\phi$ and $\psi$ are determined up to the addition by arbitrary functions of $t$, we may take without loss of generality that $b(t)=0$ at all times. 

The boundary condition in the infinite depth states that
\begin{equation}\label{E:infty}
\phi,\psi\to0\quad\text{as $y\to-\infty$}\quad\text{uniformly for $x\in\mathbb{R}$}.
\end{equation}
In the finite depth, we replace \eqref{E:infty} by 
\[
\phi_y=0\quad\text{at $y=-h$}\quad\text{for some $h>0$}
\]
and proceed in like manner \citep[see][for instance, for details]{DH1}. By the way, the effects of finite depth change the amplitude of a Stokes wave and others, but they are insignificant otherwise \citep[see][for instance, for details]{DH1}. 

\subsection{Reformulation in conformal coordinates}\label{sec:conformal}

We proceed as in \cite{DH1} and reformulate \eqref{E:Euler}, \eqref{E:surface} and \eqref{E:infty} more conveniently in conformal coordinates. In the irrotational flow setting, we refer the interested reader to \cite{DKSZ1996,ZDAV2002}, among others. In what follows, we identify $\mathbb{R}^2$ with $\mathbb{C}$ whenever it is convenient to do so. 

Suppose that 
\begin{equation}\label{def:conformal}
z(u+iv,t)=(x+iy)(u+iv,t)
\end{equation}
conformally maps $\mathbb{C}^-:=\{u+iv\in\mathbb{C}: v<0\}$ of $2\pi$ period in the $u$ variable, to $\varOmega(t)$ of $2\pi$ period in $x$, at time $t$, such that $z(u+iv)\to u+iv$ as $v\to-\infty$. Suppose that \eqref{def:conformal} extends to map $\{u+i0:u\in\mathbb{R}\}$ to $\varGamma(t)$. 

Since $z-(u+iv)$ is holomorphic in $\mathbb{C}^-$ and 
it is $2\pi$ periodic in the $u$ variable by hypothesis, a `Titchmarsh theorem' \citep[see][for instance]{Titchmarsh} implies that 
\begin{equation}\label{E:Hy}
z(u,t)=u+(\mathcal{H}+i)y(u,t)\quad\text{for $u\in\mathbb{R}$}.
\end{equation}
Therefore, 
\[
\varGamma(t)=\{(u+\mathcal{H}y(u,t),y(u,t)): u\in\mathbb{R}\}.
\] 
Here and in the sequel, $\mathcal{H}$ denotes the periodic Hilbert transform, defined by
\[
\mathcal{H}e^{iku}=-i\,\text{sgn}(k)e^{iku}\quad\text{for $k\in\mathbb{Z}$}
\]
and extended by linearity and continuity. 

Moreover, let 
\[
W(u+iv,t)=(\phi+i\psi)(z(u+iv,t),t)\quad\text{for $u+iv\in\mathbb{C}^-$}.
\]
Since $W$ is holomorphic in $\mathbb{C}^-$ by \eqref{E:phi} 
it is $2\pi$ periodic in the $u$ variable by hypothesis, 
\begin{equation}\label{E:Hphi}
W(u,t)=(1-i\mathcal{H})\phi(u,t)\quad\text{for $u\in\mathbb{R}$}.
\end{equation}

Substituting \eqref{E:Hy} and \eqref{E:Hphi} into the former equation of \eqref{E:surface}, we make an explicit calculation to arrive at 
\begin{subequations}\label{E:implicit}
\begin{equation}\label{E:implicit(a)}
(1+\mathcal{H}y_u)y_t-y_u\mathcal{H}y_t-\mathcal{H}_d\phi_u-\omega yy_u=0.
\end{equation}
Substituting \eqref{E:Hy} and \eqref{E:Hphi} into \eqref{E:bernoulli}, likewise, we make an explicit calculation to arrive at
\begin{equation}
\begin{aligned}
&((1+\mathcal{H}y_u)^2+y_u^2)(\phi_t-\omega\mathcal{H}\phi+gy-b(t))\\
&-((1+\mathcal{H}y_u)\mathcal{H}y_t+y_uy_t)\phi_u
+(y_u\mathcal{H}y_t-(1+\mathcal{H}y_u)y_t)\mathcal{H}\phi_u \\
&\hspace*{77pt}+\frac12(\phi_u^2+(\mathcal{H}\phi_u)^2)
-\omega y((1+\mathcal{H}y_u)\phi_u-y_u\mathcal{H}\phi_u)=0.
\end{aligned}
\end{equation}
\end{subequations}
They make what \cite{DH1}, for instance, derived.
Conversely, a solution of \eqref{E:implicit} gives rise to a solution of \eqref{E:Euler}, \eqref{E:surface} and \eqref{E:infty}, provided that
\begin{gather*}
u\mapsto (u+\mathcal{H}y(u,t), y(u,t)), u\in\mathbb{R}, \text{is injective}
\intertext{and}
((1+\mathcal{H}y_u)^2+y_u^2)(u,t)\neq0\quad\text{for any $u\in\mathbb{R}$}.
\end{gather*}
See \cite{DH1}, for instance, for details. 
By the way, an integration of \eqref{E:implicit(a)} over one period reveals that we may take without loss of generality that
\[
\langle y(1+\mathcal{H}y_u)\rangle=0
\]
at all times. Here and in the sequel, $\langle f \rangle$ denotes the average of a $2\pi$ periodic function $f$ over one period. 

Moreover, we solve \eqref{E:implicit} for $y_t$ and $\phi_t$ and, hence, $z_t$ and $W_t$ in the complex form. The result becomes, after an explicit calculation, 
\begin{equation}\label{E:explicit}
\begin{aligned}
z_t=&i\alpha z_u, \\
W_t=&i\alpha W_u-\beta+\frac12\omega\mathcal{P}\left((z-z^*)\left(\frac{W_u}{z_u}+\frac{W_u^*}{z_u^*}\right)\right)+i\omega W+ig(z-u)+b(t),
\end{aligned}
\end{equation}
where 
\[
\alpha=\mathcal{P}\left(\frac{W_u-W_u^*+\frac12\omega(z-z^*)(z_u-z^*_u)}{|z_u|^2}\right)
\quad\text{and}\quad \beta=\mathcal{P}\left(\frac{|W_u|^2}{|z_u|^2}\right),
\]
and $\mathcal{P}=\frac12(1-i\mathcal{H})$. 
Here and in the sequel, the asterisk denotes complex conjugation. See \cite{DTS2017}, for instance, for details. The imaginary part of the former equation of \eqref{E:explicit} and the real part of the latter make what \cite{DH1}, for instance, derived. In the irrotational flow setting, they become what \cite{DKSZ1996}, for instance, derived.

Furthermore, let
\[
R=\frac{1}{z_u}\quad\text{and}\quad V=i\frac{W_u}{z_u}.
\]
We solve \eqref{E:explicit} for $R$ and $V$, to arrive at
\begin{equation}\label{E:Dyachenko}
\begin{aligned}
R_t&=i(\alpha R_u-\alpha_uR), \\
V_t&=i(\alpha V_u-\beta_uR)+\frac12i\omega R((z-z^*)\mathcal{P}(V+V^*))_u+i\omega V+g(R-1).
\end{aligned}
\end{equation}
See \cite{DTS2017}, for instance, for details. 
In the irrotational flow setting, \eqref{E:Dyachenko} becomes the so-called Dyachenko equation \citep[see][for instance]{ZDAV2002}, which is particularly convenient for numerical solution. 

\subsection{The Stokes wave problem}\label{sec:Stokes}

We turn the attention to traveling waves of \eqref{E:explicit}, for which $z(u,t)-u$ and $W(u,t)$ are functions of $u-ct$ for some $c>0$, the speed of wave propagation, and $b(t)$ is constant. Under the assumption, we will go to a moving coordinate frame, changing $u-ct$ to $u$, whereby $t$ completely disappears. The former equation of \eqref{E:explicit} becomes
\[
\phi'=\mathcal{H}(\omega yy'+cy').
\]
Here and in the sequel, the prime denotes ordinary differentiation in the $u$ variable. The latter equation of \eqref{E:explicit}, after an explicit calculation, simplifies to
\begin{equation}\label{E:Stokes}
(c^2+2b-2gy)((1+\mathcal{H}y')^2+(y')^2)-(c+\omega y(1+\mathcal{H}y')-\omega\mathcal{H}(yy'))^2=0.
\end{equation}
See \cite{DH1}, for instance, for details. 

\subsection*{Reformulation as the modified Babenko equation}

We proceed as in \cite{DH1} and reformulate \eqref{E:Stokes} more conveniently as an equation of Babenko kind. This makes use of that $\mathcal{H}+i$ determines the boundary value of a holomorphic and $2\pi$ periodic function in $\mathbb{C}^-$ up to the addition by a real constant.

We begin by rearranging \eqref{E:Stokes} as
\begin{multline*}
(c^2+2b-2gy-\omega^2y^2)((1+\mathcal{H}y')^2+(y')^2)\\
-2\omega y(c-\omega\mathcal{H}(yy'))(1+\mathcal{H}y')=(c-\omega\mathcal{H}(yy'))^2-\omega^2y^2(y')^2.
\end{multline*}
Notice that $(c-\omega(\mathcal{H}+i)(yy'))^2$ is the boundary value of a holomorphic and $2\pi$ periodic function in $\mathbb{C}^-$, whence so is
\[
(c^2+2b-2gy-\omega^2y^2)((1+\mathcal{H}y')^2+(y')^2)-2\omega y(c-\omega\mathcal{H}(yy'))(1+\mathcal{H}y'+iy').
\]
Notice that $1/(1+\mathcal{H}y'+iy')$ is the boundary value of the holomorphic and $2\pi$ periodic function $1/z_u$ in $\mathbb{C}^-$, 
whence so is 
\[
(c^2+2b-2gy-\omega^2y^2)(1+\mathcal{H}y'-iy')-2\omega y(c-\omega\mathcal{H}(yy')).
\] 
Therefore, 
\[
(c^2+2b-2gy-\omega^2y^2)(1+\mathcal{H}y')-2\omega y(c-\omega\mathcal{H}(yy'))
=\mathcal{H}(-(c^2+2b-2gy-\omega^2y^2)y')
\]
up to the addition by a real constant. See \cite{DH1}, for instance, for details. An integration over one period reveals that
\begin{equation}\label{E:DH1}
\begin{aligned}
(c^2+2b)\mathcal{H}y'-(g+\omega c)y-g(y\mathcal{H}y'+&\mathcal{H}(yy'))\\
-\frac12\omega^2(y^2+\mathcal{H}(y^2y')+&y^2\mathcal{H}y'-2y\mathcal{H}(yy'))\\
+g\langle y&(1+\mathcal{H}y')\rangle+\omega c\langle y\rangle+\frac12\omega^2\langle y^2\rangle=0.
\end{aligned}
\end{equation}
Indeed, $\langle \mathcal{H}f'\rangle=0$ for any function $f$ and 
\[
\langle y^2\mathcal{H}y'\rangle =\frac{1}{2\pi}\int^{\pi}_{-\pi} y^2\mathcal{H}y'~du
=-\frac{1}{2\pi}\int^{\pi}_{-\pi} y\mathcal{H}(y^2)'~du=-\langle 2y\mathcal{H}(yy')\rangle.
\]
We emphasize that \eqref{E:DH1} is made up of polynomials of $y$, $y'$ and their Hilbert transforms.

But for any $\omega, c$ and $b\in\mathbb{R}$, the left side of \eqref{E:DH1} maps a $2\pi$ periodic function to a $2\pi$ periodic function of mean zero, whereas the left side of \eqref{E:Stokes} maps a $2\pi$ periodic function to a $2\pi$ periodic function, not necessarily of mean zero. 
In order to reconcile loss of information from \eqref{E:Stokes} to \eqref{E:DH1}, \cite{DH1} \citep[see also][]{CSV2016} supplemented \eqref{E:DH1} with
\begin{equation}\label{E:constraint}
\langle (c^2+2b-2gy)((1+\mathcal{H}y')^2+(y')^2)\rangle
-\langle(c+\omega y(1+\mathcal{H}y')-\omega\mathcal{H}(yy'))^2\rangle=0.
\end{equation}

But the linearization of \eqref{E:DH1}-\eqref{E:constraint} with respect to $y$ and $b$ is {\em not} self-adjoint, because the linear part of \eqref{E:DH1} in $y$ depends on $b$, among others, whence the Conjugate Gradient or Conjugate Residual method is inapplicable. \cite{DH1} employed the Generalized Minimal Residual method, instead, and achieved success. But it becomes impractical for large amplitude waves for large values of positive constant vorticity. 

Here we manipulate a change of variables
\[
y \mapsto y_0:=y+\eta_0\quad\text{and}\quad c\mapsto c_0:=c-\omega \eta_0
\]
for some $\eta_0\in\mathbb{R}$, so that $b=0$ but the linear and nonlinear parts of \eqref{E:DH1} remain invariant otherwise. This works, provided that
\[
c^2+2b-2gy=c_0^2-2gy_0.
\]
Moreover, we perform a change of variables $y\mapsto y+\eta$ for some $\eta\in\mathbb{R}$, so that the constant part of \eqref{E:DH1} vanishes. The result becomes, by abuse of notation,
\begin{equation}\label{E:Babenko}
c^2\mathcal{H}y'-(g+\omega c)y-g(y\mathcal{H}y'+\mathcal{H}(yy'))
-\frac12\omega^2(y^2+\mathcal{H}(y^2y')+y^2\mathcal{H}y'-2y\mathcal{H}(yy'))=0.
\end{equation}
An integration over one period reveals that
\[
g\langle y(1+\mathcal{H}y')\rangle+\omega c\langle y\rangle+\frac12\omega^2\langle y^2\rangle=0.
\]
Conversely, a solution of \eqref{E:DH1}-\eqref{E:constraint} and, hence, \eqref{E:Babenko} gives rise to a periodic traveling wave of \eqref{E:Euler}, \eqref{E:surface} and \eqref{E:infty}, provided that
\begin{subequations}\label{C:limiting}
\begin{equation}\label{C:touching}
u\mapsto (u+\mathcal{H}y(u), y(u)), u\in\mathbb{R}, \text{ is injective}
\end{equation}
and
\begin{equation}\label{C:extreme}
(1+\mathcal{H}y'(u))^2+y'(u)^2\neq0\quad\text{for any $u\in\mathbb{R}$}.
\end{equation}
\end{subequations}
See \cite{DH1}, for instance, for details. By the way, $u+(\mathcal{H}+i)y(u)$ makes the boundary value of a conformal mapping (see \eqref{def:conformal}). 

In what follows, we assume that $y$ is even \citep[see][for instance, for arbitrary vorticity]{Hur2007}.

To recapitulate, the Stokes wave problem in a constant vorticity flow is to seek a $2\pi$ periodic and even function $y$, satisfying \eqref{C:limiting}, which solves \eqref{E:Babenko} for some $\omega, c\in\mathbb{R}$. To compare, \cite{DH1} solved \eqref{E:DH1}-\eqref{E:constraint} for $y$ and $b$. But the associated linearized operator is not self-adjoint.

In the irrotational flow setting, \eqref{E:Babenko} becomes the well-known Babenko equation \citep[see][for instance]{Babenko1987}. One may regard \eqref{E:Babenko} as the `modified Babenko equation,' permitting constant vorticity. 

We emphasize that the linearization of \eqref{E:Babenko}:
\begin{equation}\label{def:L}
\begin{aligned}
\delta y\mapsto &c^2\mathcal{H}(\delta y)'-(g+c\omega)\delta y
-g(\delta y\mathcal{H}y'+y\mathcal{H}(\delta y)'+\mathcal{H}(y\delta y)') \\
&-\frac12\omega^2(2y\delta y+\mathcal{H}(y^2\delta y)'-[2y\delta y,y]+[y^2,\delta y]),
\end{aligned}
\end{equation}
where $[f_1,f_2]=f_1\mathcal{H}f_2-f_2\mathcal{H}f_1$, is self-adjoint, whereby we may employ the Conjugate Gradient method for efficient numerical solution. We achieve a high numerical precision for large amplitude waves for unprecedentedly large values of constant vorticity.

\section{Numerical Method}\label{sec:numerical}

\subsection{Auxiliary conformal coordinates}

\cite{DLK2016} and others discussed that a Stokes wave defines the boundary value of a conformal mapping $z=z(w)$ for $w:=u+iv\in\mathbb{C}^-$ (see \eqref{def:conformal}), and an analytical continuation of $z$ to $\mathbb{C}$ has complex singularities in $[iv_c, +i\infty)$, where $iv_c$ is the closest singularity of $z$ to $\mathbb{C}^-$; moreover,
\[
z(w)-w=\sum_{k\in\mathbb{Z},\leq0}\widehat{z}(k)e^{ikw},\quad
\text{where $|\widehat{z}(k)|\to e^{-v_c|k|}$ as $|k|\to\infty$}.
\]
In the irrotational flow setting, $v_c\to0$ as the profile tend to the wave of greatest height, or the extreme wave, presenting challenges in numerical solution. For instance, \cite{DLK2016} took up to $2^{27}\approx 1.3\times 10^8$ uniform grid points in the $u$ variable for $v_c=5.93824419\dots\times10^{-7}$. \cite{LDS2017} devised an auxiliary conformal transform, involving trigonometric functions, 
to reduce the number of (nonuniform) grid points to $\sqrt{2^{27}}\approx 4.2\times10^4$. See also \cite{DH1} in the  constant vorticity setting.

\begin{figure}
\centering
\includegraphics[scale=1]{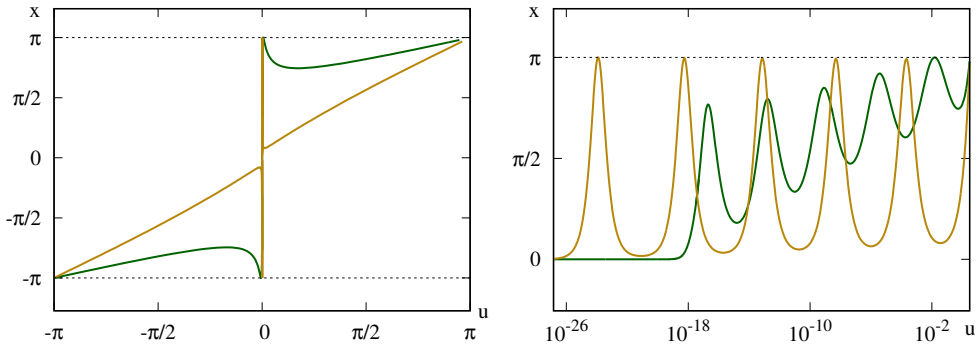}
\caption{The graphs of $x=x(u)$ for two Stokes waves, $\omega=4.0$ (green) and $\omega=14.0$ (yellow), where $(x(u),y(u))$, $u\in[-2\pi,2\pi]$, is a conformal parametrization of a wave profile: On a linear scale (left) and a logarithmic scale (right). See the last column in Figure~\ref{fig:5c} for the profiles.}
\label{fig:u-x}
\end{figure}

But for a large value of positive constant vorticity, for instance, in Figure~\ref{fig:u-x} are the graphs of $x=x(u)$ for two Stokes waves, whose profiles are in the last column of Figure~\ref{fig:5c}; $(x(u),y(u))$, $u\in[-\pi,\pi]$, makes a conformal parametrization of a profile. The right panel suggests that $v_c$ is of the order of $10^{-24}$, whence the numerical method of \cite{DLK2016, Lushnikov2016, LDS2017} would converge too slowly to be efficient. 

Here we exploit conformal mappings to vertically slit regions in $\mathbb{C}$, involving Jacobi elliptic function, for further improvements. We refer the interested reader to \cite{HT2009}, for instance, for details. For $L>0$, to be determined, let
\begin{equation}\label{def:HT}
u(q)=-\pi+2\am\left(K\frac{\pi+q}{\pi}\left|\sqrt{1-L^2}\right.\right),
\end{equation}
where $\am$ denotes the Jacobi amplitude and $K$ is the complete elliptic integral of the first kind with the elliptic modulus $\sqrt{1-L^2}$ \citep[see][for instance]{AS1964}. We extend \eqref{def:HT} to conformally map $\{q\in\mathbb{C}: -d<\text{Im}\,q<0\}$ of $2\pi$ period, to $\mathbb{C}^-$ of $2\pi$ period in the $w$ variable, where 
\[
d=\pi K'/K
\] 
and $K'$ is the complete elliptic integral of the first kind with the modulus $L$. 

An explicit calculation reveals that \eqref{def:HT} maps the Hilbert transform for the strip of depth $d$ in the $q$ variable to the Hilbert transform in the $w$ variable, so that \eqref{E:Babenko} remains invariant in the $q$ variable. See also \cite{Silantyev2018} in the irrotational flow setting. 

Moreover, \eqref{def:HT} maps uniform grid points in the $q$ variable over one period, to highly nonuniform in the $u$ variable. We wish to choose a suitable value of $L$, so that the inverse of \eqref{def:HT} maps $iv_c$, the closest singularity to $\mathbb{C}^-$ in the $u$ variable, to $\text{Im}\,q=d$. Since
\begin{equation}\label{E:d}
d \approx \frac{\pi^2}{2}\frac{1}{\log{(4/L)}} \approx \frac{\pi^2}{2}\frac{1}{\log{(8/v_c)}}
\quad\text{as $v_c\to 0$}, 
\end{equation}
it will enlarge considerably the region of analyticity in the $q$ variable when $v_c$ is small. For instance, if the closest singularity to $\mathbb{C}^-$ is of the order of $10^{-24}$ in the $u$ variable, the corresponding singularity is about $0.086$ in the $q$ variable, and in practice, it necessitates no more than $4096$ uniform grid points in the $q$ variable to resolve a numerical solution in quadruple precision.

\begin{remark*}\rm
Numerical evaluation of Jacobi elliptic functions in \eqref{def:HT} is prone to catastrophic loss of precision in finite digit arithmetic, and it becomes more dramatic when $L$ decreases. Here we employ Landen transformations to evaluate the Jacobi amplitude, and the GNU MPFR library in arbitrary precision floating point arithmetic. 
We then truncate the result to quadruple precision arithmetic. We employ the arithmetic--geometric mean method to compute complete elliptic integrals. We refer the interested reader to \cite{PTVF, Bulirsch1965}, for instance, for details.
\end{remark*}

\begin{remark*}\rm
We use \eqref{E:d} to approximate the optimal value of $L$, but the exact value of $iv_c$, the closest singularity to $\mathbb{C}^-$ in the $u$ variable, is by and large unknown. When $L$ is smaller than optimal then the inverse of \eqref{def:HT} maps $iv_c$ to $\pm a+id$ for some $a\in\mathbb{R}$, and $d$ depends on $L$ logarithmically. But if $L$ is larger than optimal then $iv_c$ is no longer mapped to $\text{Im}\,q=d$, and the Fourier spectrum in the $q$ variable widens rapidly.
\end{remark*}

\subsection{The Newton-CG method}

We write \eqref{E:Babenko} in the operator form as 
\begin{equation}\label{E:G=0}
\mathcal{G}(y, c, \omega)=0,
\end{equation}
where $y=y(u(q))$ and $q$ is in \eqref{def:HT}, and we solve it iteratively using the Newton method. Let
\[
y^{(n+1)}=y^{(n)}+\delta y^{(n)}\quad \text{for $n=0,1,2,\dots$},
\] 
where $y^{(0)}$ is an initial guess, to be supplied, and $\delta y^{(n)}$ solves 
\begin{equation}\label{E:dG}
\delta\mathcal{G}(y^{(n)},c,\omega)\delta y^{(n)}=-\mathcal{G}(y^{(n)},c,\omega),
\end{equation}
$\delta\mathcal{G}(y^{(n)},c,\omega)$ is the linearization of $\mathcal{G}(y,c,\omega)$ with respect to $y$ and evaluated at $y=y^{(n)}$ (see \eqref{def:L}). 

We take uniform grid points in the $q$ variable over one period. We approximate $y^{(n)}$ by a discrete Fourier transform and numerically evaluate using a fast Fourier transform. We numerically approximate polynomials of $y^{(n)}$, $(y^{(n)})'$ and their Hilbert transforms likewise and, hence, $\mathcal{G}(y^{(n)},c,\omega)$ and $\delta\mathcal{G}(y^{(n)},c,\omega)$. 

We numerically solve \eqref{E:dG} using the Conjugate Gradient (CG) method \citep[see][for instance]{Yang2009, Yang2010}, in the Fourier space for the $q$ variable. Here we choose the linear part of $\mathcal{G}(y^{(n)},c,\omega)$ as a preconditioner. See \cite{DLK2016}, for instance, for details. We merely pause to remark that $\delta\mathcal{G}(y^{(n)},c,\omega)$ may not be positive definite, but the CG method converges nonetheless. To compare, the linearization of \eqref{E:DH1}-\eqref{E:constraint} is {\em not} self-adjoint, whence the CG method is inapplicable. \cite{DH1} employed the Generalized Minimal Residual Method \citep[see][for instance]{Saad2003}, instead, and achieved some success. 
The CG method is more powerful than the Generalized Minimal Residual method for a self-adjoint problem, and we achieve a high numerical precision for large amplitude waves for unprecedentedly large values of constant vorticity.

It turns out that we must supply an initial guess sufficiently close to a true solution of \eqref{E:G=0}, so that the Newton-CG method will converge. We begin by taking $\omega=0$ and a numerical solution of small amplitude \citep[see][for instance]{DLK2016}, and we continue the solution along in $\omega$, and $c$, if necessary, by taking the prior convergent solution as the initial guess and solving \eqref{E:G=0}, until we arrive at a solution of small amplitude for the desired $\omega$. See \cite{DH1}, for instance, for details. We then fix $\omega$ and continue the numerical solution along in $c$. 

But the results in Section~\ref{sec:result} \citep[see also][]{DH1} reveal that for $\omega$ positive and large, a gap of `unphysical' solutions appears in the wave speed versus amplitude plane; moreover, the gap becomes abnormally large in size as $\omega$ increases, so that it is impractical or even numerically prohibited to continue along in $c$ to trace out the gap. For instance, Figure~\ref{fig:c-s} indicates that for $\omega\geq2.4$, the wave speed reaches the order of hundreds in the gap. We then take a smaller $\omega$ and a physical numerical solution and we continue along in $\omega$, and $c$, if necessary, until we arrive at a physical solution for the desired $\omega$. See \cite{DH1}, for instance, for details. 

The results herein are carried out in quadruple precision floating point arithmetic. 
We say that the Newton method converges if the residuals $\lesssim N^{1/2}10^{-28}$, where $N$ is the number of uniform grid points in the $q$ variable. 
We take $N$ sufficiently large, so that the Fourier coefficient $\lesssim N^{-1/2}10^{-26}$ at the wave number $N$, to guarantee that the error in approximating a numerical solution by a discrete Fourier transform is inconsequential.
To compare, \cite{DH1} required the residuals less than $10^{-13}$ and the Fourier coefficients less thans $10^{-12}$.

\section{Results}\label{sec:result}

Here we assume without loss of generality that $c$ is positive, and we permit $\omega$ positive or negative. Indeed, if $y$ and $c$ solve \eqref{E:Babenko} for some $\omega$ then so do $y$ and $-c$ for $-\omega$. We take for simplicity that $g=1$. 

In what follows, $s$ denotes the steepness, the crest-to-trough wave height divided by the period $=2\pi$, and $c$ measures the wave speed, for which $\langle y(1+\mathcal{H}y')\rangle=0$. This is what \cite{SS1985,PTdS1988,DH1} chose, among others.

\subsection{Folds and gaps}

\begin{figure}
\centerline{
\includegraphics{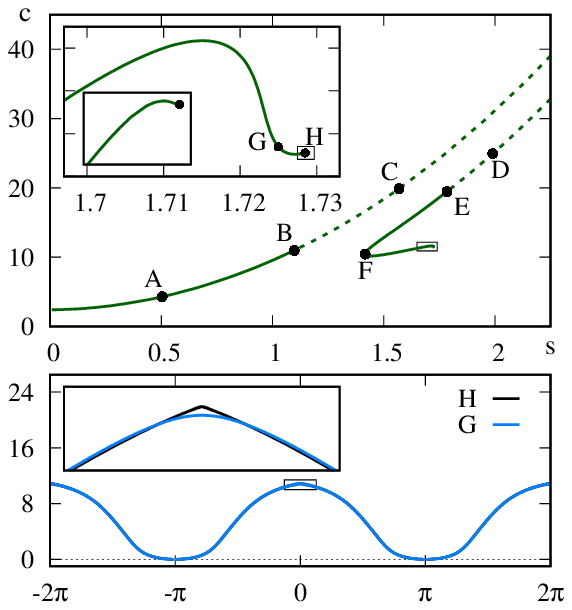}  
\includegraphics{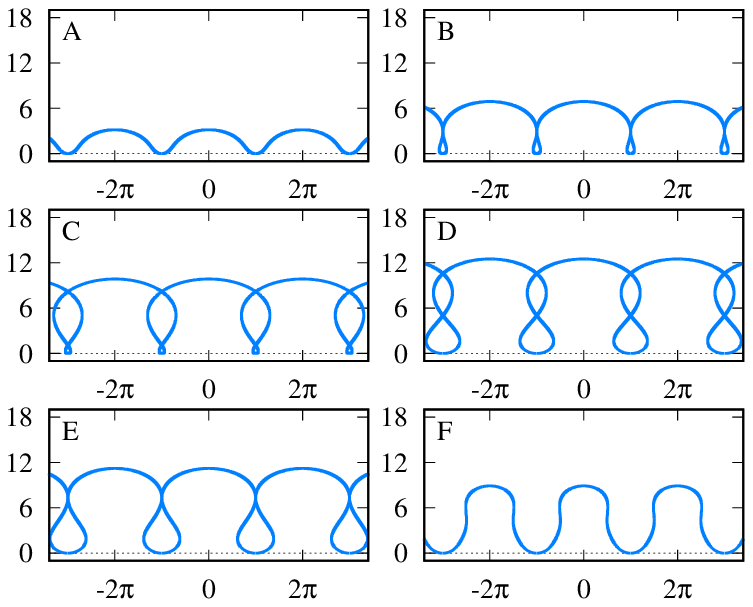}}
\caption{$\omega=2$: On the upper left, wave speed versus steepness. Solid curves are for physical solutions and dashed curves for unphysical. The insets are closeups near the endpoint of the continuation of the numerical solution. On the right, the wave profiles of six solutions, in the $(x,y)$ plane in the range $x\in[-4\pi,4\pi]$, labelled by $A$ through $F$ in the $c=c(s)$ curve. Troughs are at $y=0$. On the lower left, the profiles of two almost extreme waves, in the $(x,y)$ plane in the range $x\in[-2\pi,2\pi]$, labelled by $G$ and $H$ in the $c=c(s)$ curve. The inset is a closeup in the vicinity of the crests.}
\label{fig:w2}
\end{figure}

For a large value of positive constant vorticity, \cite{SS1985} discovered, and \cite{DH1} rediscovered, that the steepness increases, decreases and increases during the continuation of the numerical solution of the Stokes wave problem. Namely, a fold appears in the wave speed versus steepness plane; moreover, as the value of the vorticity increases, unphysical solutions make a gap in the wave speed versus steepness plane. 

For instance, for $\omega=2.0$, the upper left panel of Figure~\ref{fig:w2} collects the wave speed versus steepness from the continuation of the numerical solution of \eqref{E:Babenko}. The right and lower left panels display, in the $(x,y)$ plane, the wave profiles of eight solutions, labelled by $A$ through $H$ along the $c=c(s)$ curve, in the order in which they appear in the numerical continuation. Here the troughs are at $y=0$. This is in excellent agreement with \citep[][Figure~10 and Figure~11]{SS1985}, qualitatively and quantitatively. A caveat is that the vorticity of \cite{SS1985} differs in sign. We refer the interested reader to \cite{SS1985, DH1}, among others, for $\omega$ negative, or positive but small.

The upper left panel reveals a fold in the $c=c(s)$ curve, along which $s$ increases and decreases. For $s$ small, for instance, for wave $A$, the profile is the graph of a single valued function. But we observe that the profile becomes more rounded as $s$ increases along the fold, so that overhanging waves appear, whose profile is no longer the graph of a single valued function; moreover, a touching wave follows, whose profile contacts with itself at the trough line, enclosing a bubble of air. For instance, wave $B$ is an almost touching wave. By the way, steep and rounded waves have relevance to backwash at the beach \citep[see][for instance]{PTdS1988}. 

We report that a numerical solution turns unphysical past the touching wave because \eqref{C:touching} no longer holds true. For instance, for wave $C$, the profile intersects itself and the fluid region overlaps itself. We observe that the wave profile becomes less rounded as $s$ decreases along the fold, so that another touching wave appears; moreover, a numerical solution turns physical past the touching wave. For instance, wave $D$ is unphysical, wave $E$ is an almost touching wave, and wave $F$ makes a physical solution past the touching wave. Together, a gap exists in the $c=c(s)$ curve, which consists of unphysical numerical solutions and which is bounded by two touching waves. 

We merely pause to remark that the numerical method of \cite{SS1985} and others diverges in a gap. In stark contrast, the numerical method herein converges throughout. By the way, we discontinue a unphysical numerical solution when $c$ becomes abnormally large. Instead, we take a smaller $\omega$ and a physical numerical solution and we continue the solution along in $\omega$, and $c$, if necessary. See \cite{DH1}, for instance, for details.

The insets of the upper left panel suggest that $s$ increases monotonically past the end of the fold, while $c$ experiences oscillations, like in the irrotational flow setting \citep[see][for instance]{DLK2016, LDS2017}. The lower left panel suggests that overhanging waves disappear as $s$ increases past the end of the fold; moreover, the inset reveals that the crests become sharper, like when $\omega=0$. We claim that an extreme wave appears ultimately, which is distinguished by peaking at the crest. One may not continue the numerical solution past the extreme wave because \eqref{C:extreme} no longer holds true. For instance, wave $H$ is an almost extreme wave, and we calculate $s=1.72849105$. It is noticeably higher than the celebrated wave of greatest height or the extreme wave in the irrotational setting \citep[see][for instance]{DLK2016}. But the extreme wave here seems no longer the wave of greatest height. For instance, for wave~$E$, we calculate $s=1.78406376$. By the way, \cite{SS1985} estimated that the extreme wave steepness~$\approx1.727$. The method herein offers a significant improvement.

\begin{figure}
\centerline{
\includegraphics{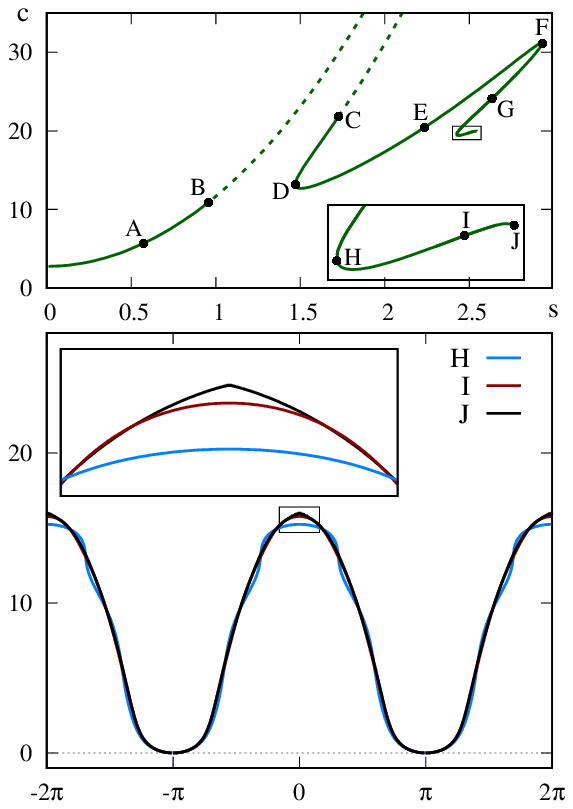}
\includegraphics{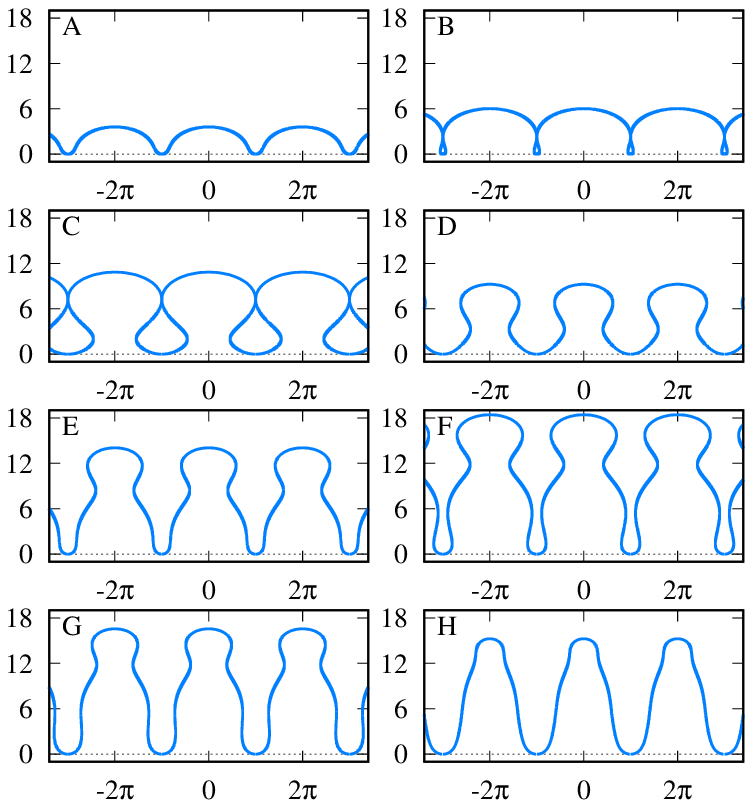}}
\caption{$\omega=2.4$: On the upper left, wave speed versus steepness. Solid and dashed curves are for physical and unphysical solutions, respectively. The inset is a closeup near the endpoint of the continuation of the numerical solution. On the right, the wave profiles of eight solutions, in the $(x,y)$ plane in the range $x\in[-4\pi,4\pi]$, labelled by $A$ through $H$ in the $c=c(s)$ curve. Troughs are at $y=0$. On the lower left, the profiles of three almost extreme waves, in the $(x,y)$ plane in the range $x\in[-2\pi,2\pi]$, labelled by $H$ through $J$ in the $c=c(s)$ curve. The inset is a closeup in the vicinity of the crests.}
\label{fig:w2.4}
\end{figure}

\begin{figure}
\centerline{
\includegraphics{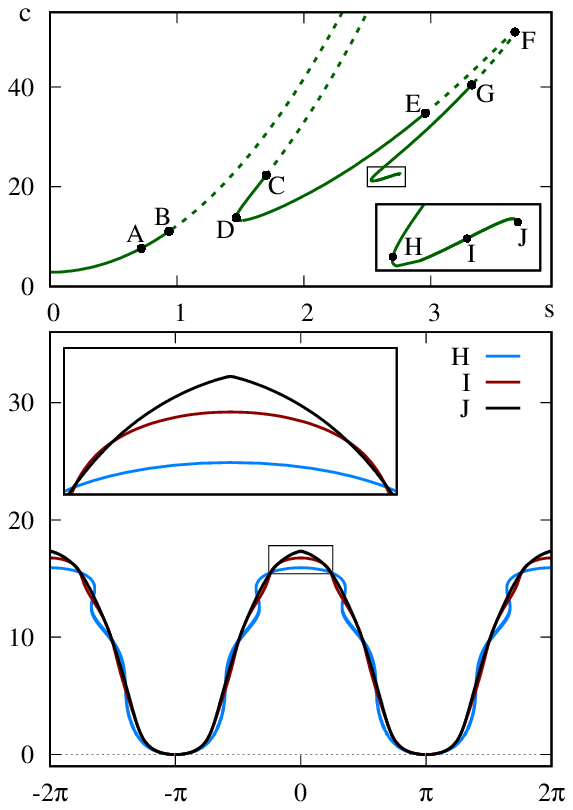}
\includegraphics{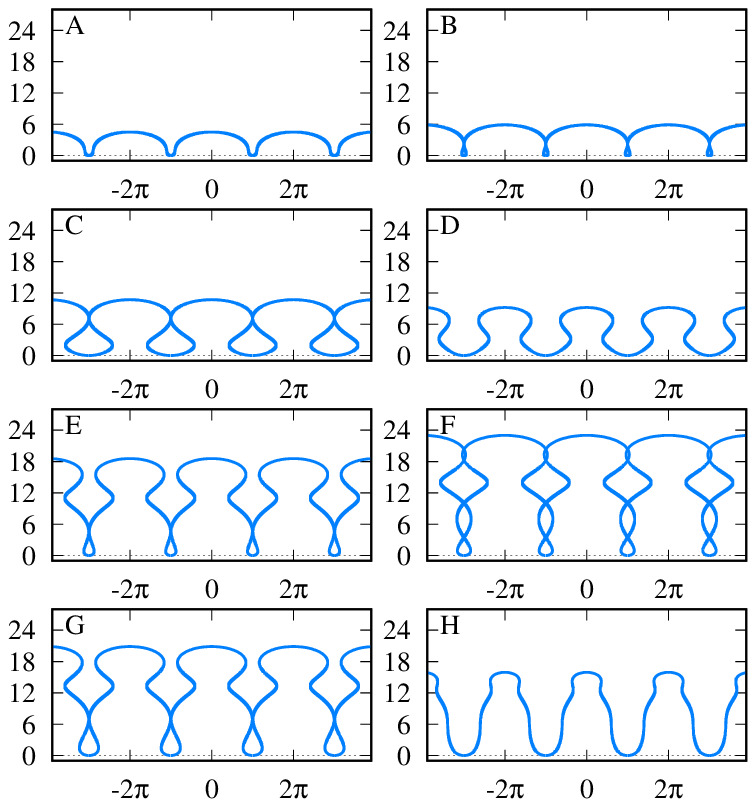}}
\caption{$\omega=2.5$: On the upper left, wave speed versus steepness. Solid and dashed curves are for physical and unphysical solutions. The inset is a closeup near the endpoint of the continuation of the numerical solution. On the right, the wave profiles of eight solutions, labelled by $A$ through $H$ in the $c=c(s)$ curve. On the lower left, the profiles of three almost extreme waves, labelled by $H$ through $J$ in the $c=c(s)$ curve. The inset is a closeup in the vicinity of the crests.}
\label{fig:w2.5}
\end{figure}

For a larger value of the positive constant vorticity, we take matters further and discover a second fold and a second gap in the wave speed versus steepness plane!

For instance, for $\omega=2.4$, in the upper left panel osf Figure~\ref{fig:w2.4} is the $c=c(s)$ curve from the continuation of the numerical solution of \eqref{E:Babenko}, and in the right and the lower left panels are the wave profiles of ten solutions at the indicated points along the $c=c(s)$ curve. The upper left panel reveals a fold, in the range $s=0$ to wave $D$, and a gap, bounded by waves $B$ and $C$, like Figure~\ref{fig:w2}. The profile of wave~$A$ is the graph of a single valued function. Wave $B$ is an almost touching wave near the beginning of the gap, and overhanging waves appear somewhere waves $A$ to $B$. Wave $C$ is an almost touching wave near the end of the gap. A numerical solution in the gap becomes unphysical because \eqref{C:touching} no longer holds true. Wave $D$ is near the end of the fold. 
 
Moreover, the upper left panel reveals that $s$ increases from waves $D$ to $F$, decreases from waves $F$ to $H$, and increases past wave $H$, past the end of the fold. Together, another fold exists in the $c=c(s)$ curve. 

The inset of the upper left panel suggests that $s$ increases monotonically past the end of the second fold, while $c$ experiences oscillations, like Figure~\ref{fig:w2} past the end of the fold. The lower left panel suggests that overhanging waves disappear as $s$ increases past the end of the second fold; moreover, the inset reveals that the crests become sharper, like when $\omega=2.0$. Therefore, we claim that an extreme wave appears ultimately as $s$ increases past the end of the second fold, and one may not continue the numerical solution past the extreme wave, because \eqref{C:extreme} no longer holds true. For instance, wave $J$ is an almost extreme wave.

For $\omega=2.5$, in Figure~\ref{fig:w2.5} are the $c=c(s)$ curve and a selection of wave profiles at the indicated points along the $c=c(s)$ curve. The upper left panel reveals the lowest fold, in the range $s=0$ to wave $D$, and a gap, bounded by waves $B$ to $C$, and the second fold, from waves $D$ to $F$, like Figure~\ref{fig:w2.4}.

We observe that the wave profile becomes more rounded as $s$ increases along the second fold, so that a touching wave appears, like along the lowest fold. For instance, wave $E$ is an almost touching wave. We observe that a numerical solution turns unphysical past the touching wave. For instance, for wave $F$, the profile intersects itself and the fluid region overlaps itself. We observe that the wave profile becomes less rounded as $s$ decreases along the second fold, so that another touching wave appears; moreover, a numerical solution turns physical past the touching wave, like along the lowest fold. For instance, wave $G$ is an almost touching wave and wave $H$ makes a physical solution past the touching wave. Together, a second gap exists in the second fold, consisting of unphysical numerical solutions and bounded by two touching waves. 

We merely pause to remark that the numerical method of \cite{SS1985} and others diverges in a gap. In order to seek physical numerical solutions past the gap, \cite{SS1985} took $\omega=0$ and an almost extreme wave, manipulating that \eqref{C:extreme} no longer holds true, and they continued the numerical solution along in $\omega$, and $c$, if necessary, until they arrived at an almost extreme wave for the desired $\omega$. They then continued the numerical solution along in $c$ until they arrived at an almost touching wave, past which their numerical method would diverge. This works, provided that there is one gap, but it is incapable of finding a second gap. To the best of the authors' knowledge, a second gap reported herein is novel! By the way, the numerical method of \cite{DH1} converges in a gap, but it would take too much time to accurately resolve a numerical solution along a second fold.

The upper left panel suggests that $s$ increases monotonically past the second fold, while $c$ experiences oscillations, like Figure~\ref{fig:w2.4}; moreover an extreme wave appears ultimately as $s$ increases past the second fold. For instance, wsave $J$ is an almost extreme wave.

\begin{figure}
\centerline{
\includegraphics{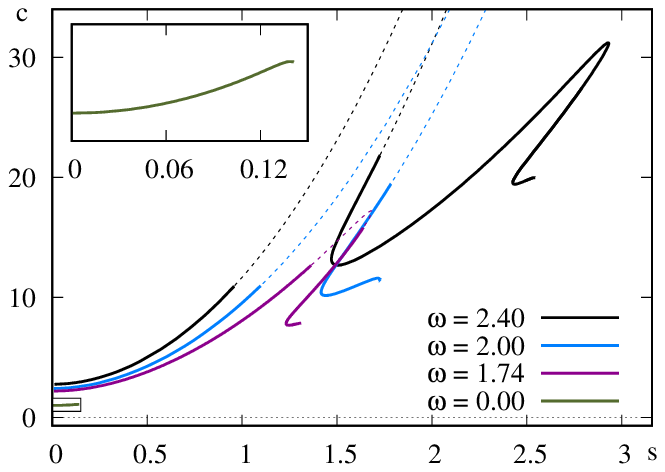}
\includegraphics{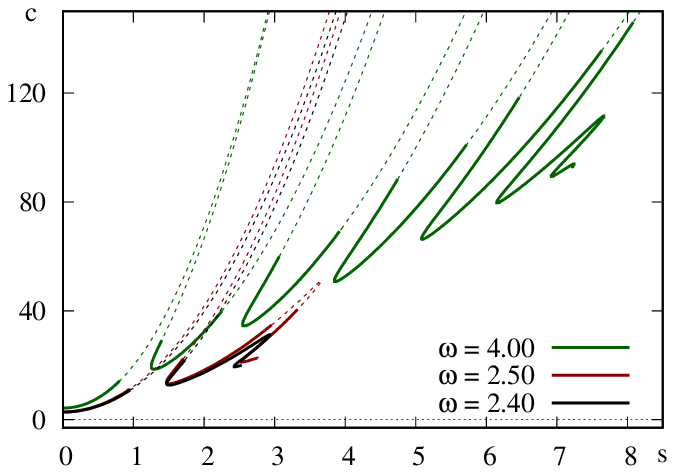}}
\caption{Wave speed versus steepness: On  the left, for $\omega=0$~(green), $1.74$~(magenta), $2.0$~(blue), $2.4$~(black) and on the right, for $\omega=2.4$~(black), $2.5$~(red), $4.0$~(green). Solid curves are for physical solutions and dashed curves for unphysical. The inset is a closeup for $\omega=0$.}
\label{fig:c-s}
\end{figure}

\begin{figure}
\centerline{
\includegraphics[scale=1]{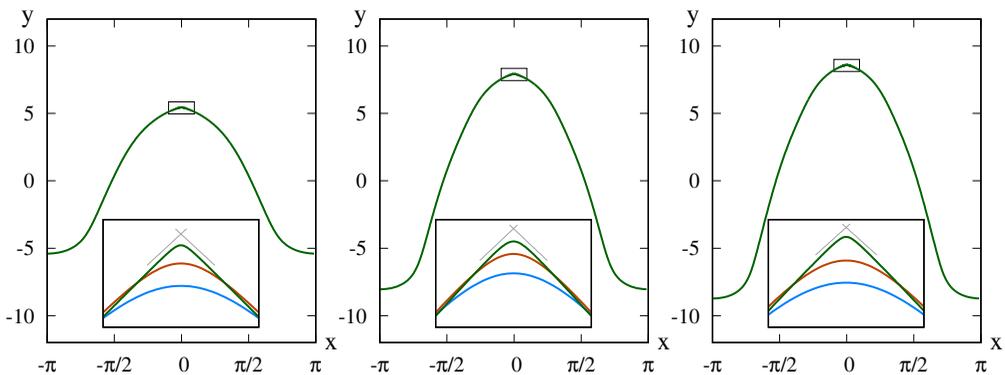}}
\caption{Almost extreme waves: $\omega=2.0$ (left), $\omega=2.4$ (center) and $\omega=2.5$ (right), in the $(x,y)$ plane in the range $x\in[-\pi,\pi]$. The mean fluid level is at $y=0$. The insets are closeups in the vicinities of the crests. The asymptotics make a $120^\circ$ angle.}
\label{fig:extreme}
\end{figure}

To recapitulate, Figure~\ref{fig:c-s} includes the wave speed versus steepness for several values of positive constant vorticity. For zero vorticity, \cite{LHF1978} and others predicted that the wave speed experiences infinitely many oscillations while the steepness increases monotonically toward the wave of greatest height. Numerical computations \citep[see][for instance, and references therein]{DLK2016, LDS2017} bear it out. The inset of the left panel reproduces it. For negative constant vorticity, the crests become sharper and lower \citep[see][for instance]{SS1985,PTdS1988, DH1}. For a large value of positive constant vorticity, on the other hand, we find that the lowest oscillation of the wave speed deforms into a fold and, as the value of the vorticity increases, part of the fold transforms into a gap. For instance, for $\omega=1.74$, the left panel displays a fold and a gap. We refer the interested reader to \cite{SS1985,DH1}, for instance, for details. We observe that the lowest fold and the lowest gap become larger in size as $\omega$ increases. For instance, for $\omega=2.0$, the fold and the gap are larger than $\omega=1.74$. 

Moreover, for a larger value of the vorticity, for instance, for $\omega=2.4$, the figure reveals that the second oscillation of the wave speed deforms into another fold. As the value of the vorticity increases, for instance, for $\omega=2.5$, the right panel reveals that part of the second fold transforms into another gap. We observe that the second fold and the second gap become larger in size as $\omega$ increases. Furthermore, we find that higher folds and higher gaps develop in like manner. For instance, for $\omega=4.0$, the right panel reveals five folds and five gaps!

We observe that the steepness increases monotonically past all folds, although the wave speed experiences oscillations; moreover, overhanging waves disappear as $s$ increases past all folds, and the crests become sharper, like in the irrotational setting. Therefore, we claim that an extreme wave appears ultimately, which is distinguished by peaking at the crest. For instance, Figure~\ref{fig:extreme} includes almost extreme waves for three values of positive constant vorticity, and bears it out. We calculate $\omega=2.0$, $s=1.72914844$; $\omega=2.4$, $s=2.54357018$, and $\omega=2.5$, $s=2.75939536$. Moreover, the figure suggests the extreme wave exhibits a $120^\circ$ peaking at the crest for any constant vorticity. 

\subsection{Touching waves in the infinite vorticity limit}

\begin{figure}
\centerline{
\includegraphics[scale=1]{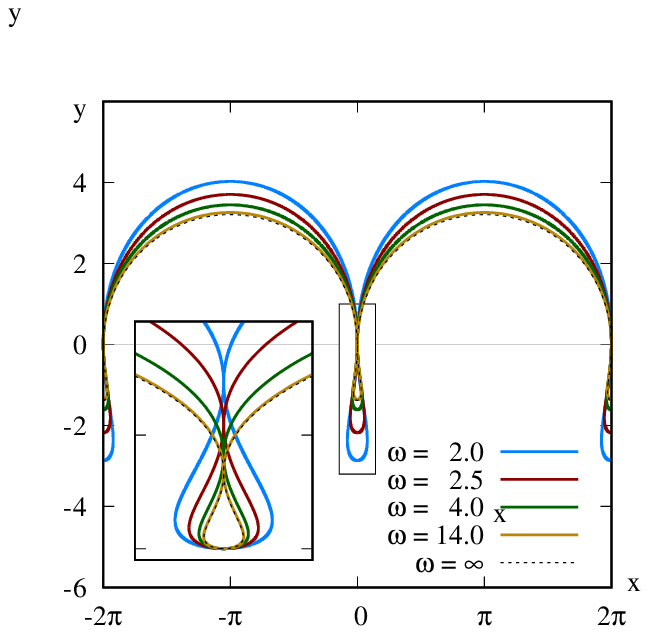}
\includegraphics[scale=1]{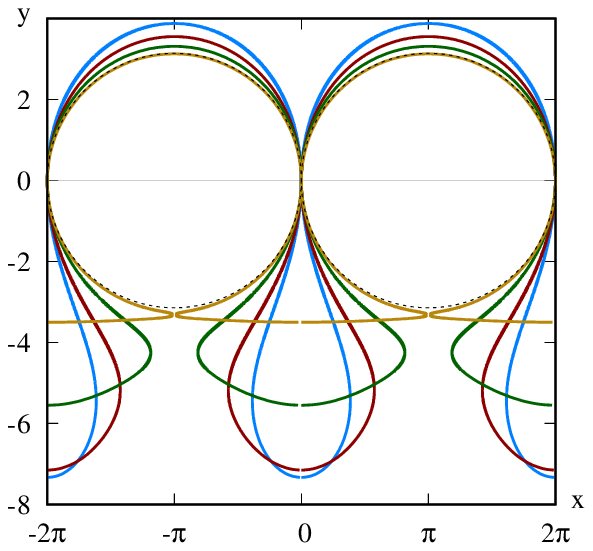}}
\caption{On the left, the profiles of almost touching waves near the beginnings of the lowest gaps for $\omega=2.0$ (blue), $2.5$ (red), $4.0$ (green) and $14.0$ (yellow), in the $(x,y)$ plane in the range $x\in[-2\pi,2\pi]$. The dashed curve represents the limiting Crapper wave in the infinite vorticity limit. The inset is a closeup in the vicinity of bubbles of air. On the right, almost touching waves near the ends of the gaps. The dashed curve represents a circle in the infinite vorticity limit. Touching is at $y=0$.}
\label{fig:gap1}
\end{figure}

Interestingly, \cite{DH1} discovered that for a sufficiently large value of positive constant vorticity, the touching waves at the boundaries of the lowest gap have capillary effects. 

The left panel of Figure~\ref{fig:gap1} displays the profiles of almost touching waves near the beginnings of the lowest gaps for four values of positive constant vorticity. We calculate $\omega=2.0$, $s=1.09717902$; $\omega=2.5$, $s=0.93652573$;  $\omega=4.0$, $s=0.80492141$, and $\omega=14.0$, $s=0.73575899$. Indeed, numerical computations \citep[see][for instance]{DH1} suggest that $s$ decreases monotonically toward $\approx0.73$ as $\omega \to \infty$. 

In the irrotational setting, \cite{Crapper1957} produced an exact solution of the capillary wave problem (in the absence of gravitational effects). He deduced that the crests become more rounded as the amplitude increases, opposite to gravity waves, and that the wave of greatest height is distinguished by the profile of a touching wave, for which $s\approx0.730$. Moreover, the left panel reveals that for instance, for $\omega=14.0$, the profile of an almost touching wave matches remarkably well the limiting Crapper wave. Therefore, we claim that touching waves at the beginnings of the lowest gaps tend to the limiting Crapper wave as the value of positive constant vorticity increases unboundedly. This reveals a striking and surprising link between rotational and capillary effects.

In the right panel of Figure~\ref{fig:gap1} are the profiles of almost touching waves near the ends of the lowest gaps for four values of positive constant vorticity. We calculate $\omega=2.0$, $s=1.78406173$; $\omega=2.5$, $s=1.70366636$; $\omega=4.0$, $s=1.41081893$, and $\omega = 14.0$, $s=1.05512149$. Indeed, numerical computations \citep[see][for instance]{DH1} suggest that $s\to1$ as $\omega\to\infty$. \cite{PTdS1988} numerically computed `pure rotational waves' in a constant vorticity flow, in the absence of gravitational effects, and argued that a limiting wave is distinguished by a circular configuration of fluid in rigid body rotation. Moreover, the right panel reveals that for instance, for $\omega=14.0$, the profile of an almost touching wave is nearly circular. Therefore, we claim that touching waves at the ends of the lowest gaps tend to the `circular vortex wave' as $\omega\to\infty$. 

\begin{figure}
\centerline{
\includegraphics[scale=1]{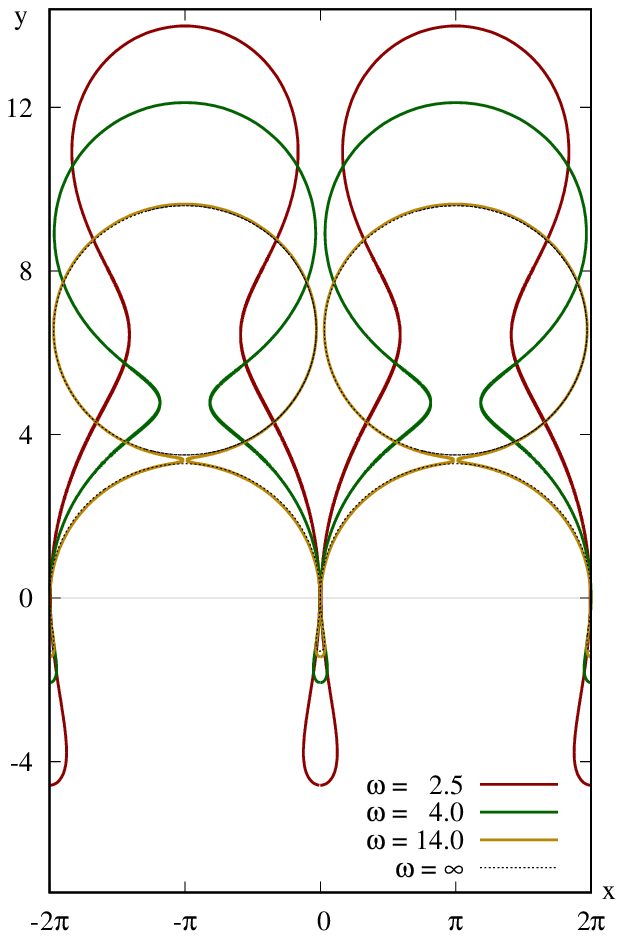}
\includegraphics[scale=1]{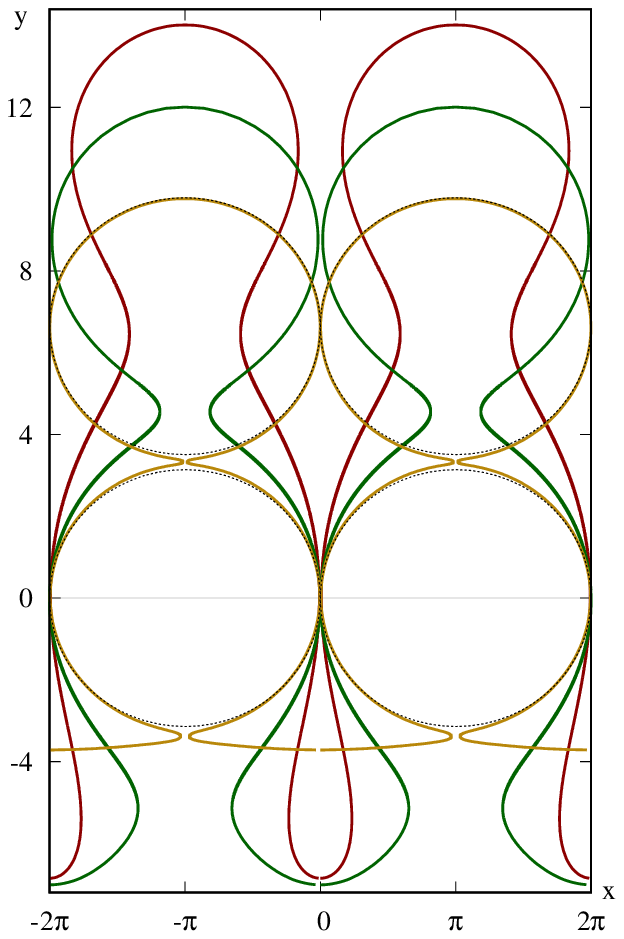}}
\caption{On the left, the profiles of almost touching waves near the beginnings of the second gaps for $\omega=2.5$ 
(red), $4.0$ (green) and $14.0$ (yellow), in the $(x,y)$ plane in the range $[-2\pi,2\pi]$. The dashed curve represents a circle on top of the limiting Crapper wave in the infinite vorticity limit. On the right, almost touching waves near the ends of the gaps. The dashed curve represents a circle on top of itself.}
\label{fig:gap2}
\end{figure}

Moreover, in the left panel of Figure~\ref{fig:gap2} are the profiles of almost touching waves near the beginnings of the second gaps, and the right panel near the ends of the gaps, for three values of positive constant vorticity. They lead to that touching waves at the beginnings of the second gaps tend to the circular vortex wave on top of the limiting Crapper wave as the value of positive constant vorticity increases unboundedly, while the circular vortex wave on top of itself at the end of the gaps in the infinite vorticity limit.

\begin{figure}
\centerline{\includegraphics[scale=1]{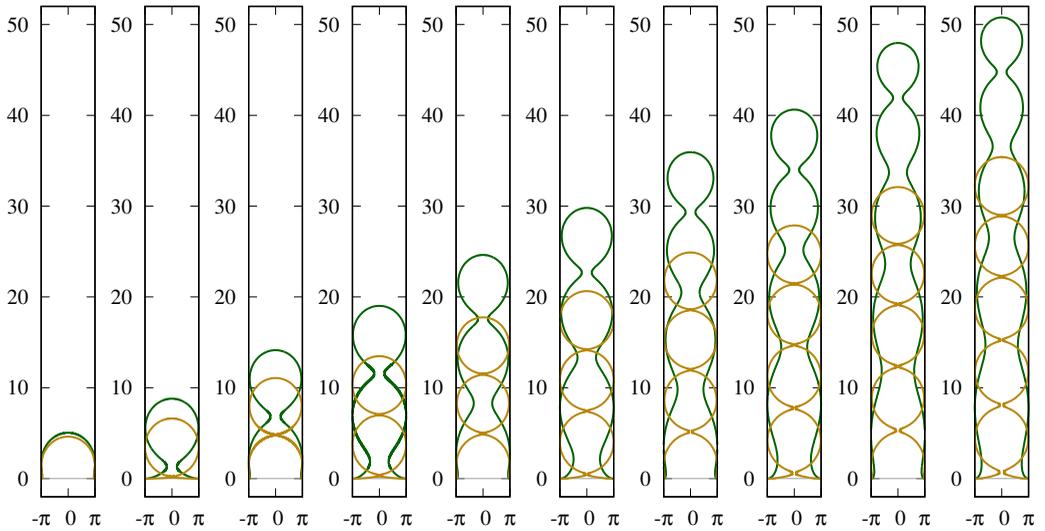}}
\caption{The profiles of almost touching waves at the beginnings and ends of the five gaps for $\omega=4.0$~(green) and $\omega=14$~(yellow), respectively, in the $(x,y)$ plane in the range $x\in[-\pi,\pi]$.}
\label{fig:5c}
\end{figure}

We claim that touching waves at the boundaries of higher gaps accommodate more circular bubbles of fluid in like manner! For instance, Figure~\ref{fig:5c} displays almost touching waves near the beginnings and ends from the lowest to fifth gaps for two large values of positive constant vorticity, and suggests that touching waves at the beginnings of the $n$-th gaps tend to $n-1$ circular vortex waves on top of the limiting Crapper wave, while $n$ circular vortex waves at the ends of the $n$-th gaps as $\omega\to\infty$. 

\subsection*{Acknowledgements}
VMH is supported by the National Science Foundation under the Faculty Early Career Development (CAREER) Award DMS-1352597. SD is supported by the National Science Foundation under DMS-1716822.

\bibliographystyle{jfm}
\bibliography{vorticitybib}

\end{document}